\documentclass[times]{elsarticle}

\usepackage{tikz}
\usepackage{pgfplots}
\usepackage{verbatim}
\usepackage{amsmath,amsfonts,amsbsy,amssymb}
\usepackage{bm}
\newcommand{\dt}[1]{\frac{d #1}{dt}} 
\newcommand{\pd}[2]{\frac{\partial #1}{\partial #2}} 
\newcommand{\pdt}[1]{\frac{\partial #1}{\partial t}}

\newcommand{\grad}{\mathop{\rm grad}\nolimits}
\renewcommand{\div}{\mathop{\rm div}\nolimits}

\journal{arXiv.org}

\newtheorem{thm}{Theorem}
\newtheorem{lm}{Lemma}
\newproof{pf}{Proof}

\begin{document}

\begin{frontmatter}

\title{Splitting schemes with respect to physical processes for double-porosity poroelasticity problems\tnoteref{label1}}
\tnotetext[label1]{This work was supported by the Russian Foundation for Basic Research (projects 14-01-00785).}

\author{A.E. Kolesov\fnref{lab2}}
\ead{kolesov.svfu@gmail.com}

\author{P.N. Vabishchevich\corref{cor1}\fnref{lab1}}
\ead{vabishchevich@gmail.com}
\cortext[cor1]{Correspondibg author.}

\address[lab2]{North-Eastern Federal University,
	      58, Belinskogo, 677000 Yakutsk, Russia}

\address[lab1]{Nuclear Safety Institute, Russian Academy of Sciences,
              52, B. Tulskaya, 115191 Moscow, Russia}

\begin{abstract}

We consider unsteady poroelasticity problem in fractured porous medium within the classical Barenblatt double-porosity model.
For numerical solution of double-porosity poroelasticity problems we construct splitting schemes with respect to physical processes, where transition to a new time level is associated with solving separate problem for the displacements and fluid pressures in pores and fractures. 
The stability of schemes is achieved by switching to three-level explicit-implicit difference scheme with some of the terms in the system of equations taken from the lower time level and by choosing a weight parameter used as a regularization parameter. The computational algorithm is based on the finite element approximation in space. The investigation of stability of splitting schemes is based on the general stability (well-posedness) theory of operator-difference schemes.
A priori estimates for proposed splitting schemes and the standard two-level scheme are provided.  The accuracy and stability of considered schemes are demonstrated by numerical experiments.

\end{abstract}


\end{frontmatter}

\section{Introduction}

Some aquifer and reservoir systems are formed by fractured rocks, which contain both pores and interconnected fractures. The fluid flow in such porous medium  is most frequently modeled by the Barenblatt double-porosity model \cite{Bai1993,Barenblatt1960}, where we consider two diffusion processes coupled by an exchange term. 
Also, the pressures of fluid in both pores and fractures contribute to deformation of porous medium. Therefore, we incorporate Biot poroelasticity model \cite{Biot1941} into Barenblatt model. 

The mathematical model of double-porosity poroelasticity problem includes the Lame equation for the displacement and  two nonstationary parabolic equations for the fluid pressures in pores and fractures \cite{Beskos1986,Wilson1982}.
Note that the displacement equation contains body forces, which are proportional to the gradient of pressures.
In turn, the pressure equations includes dilation terms described by the divergence of the displacement velocity.

The numerical solution of poroelasticity problems is usually based on a finite element approximation in space \cite{Phillips2008a,Wheeler2013}. It is well-known that in order to satisfy LBB-condition (inf-sup condition) \cite{Brezzi1991} we must use mixed finite elements, where the order of approximation for pressure is lower then the order of approximation for displacement. The finite element formulation of poroelasticity problem with double-porosity is presented in \cite{khaled1984}. 

For the time approximation the standard two-level finite difference schemes with weights are used. 
In \cite{Boal2012a} the stability and convergence of two-level schemes for double-porosity poroelasticity problems with fi\-nite-dif\-ference approximations in space were analyzed using the general stability (correctness) theory of operator-difference schemes \cite{Samarskii1989,Samarskii2002}. 
The numerical implementation of two-level schemes is associated with solving coupled system of equations, which requires the use of special algorithms \cite{Alexsson2013,Gaspar2007a} or additive operator-difference schemes (splitting schemes) \cite{Marchuk1990,Vabishchevich2014}.

Splitting schemes are designed for the efficient computational realization of various unsteady problems by switching to a chain of simpler problems. 
The simplest example of such schemes is splitting with respect to spatial variables (alternating direction methods). Regionally additive  schemes or domain decomposition methods are widely used for computations on parallel computers.
For poroelasticity problems we use splitting with respect to physical processes, when the transition to a new time level is performed by sequentially solving separate problems for displacement and pressure.

Number of splitting schemes for poroelasticity problems are constructed and used in \cite{Armero1992,bukavc2015partitioning,bukavc2015operator,Jha2007,Kim2011a,Kim2011b,Kolesov2014,Mikelic2012,Wheeler2007}.
The analysis of stability and performance of these schemes is usually conducted using methodical computations without a theoretical study. 
Rigorous mathematical results concerning the stability of splitting sche\-mes for poroelasticity problems are described in \cite{Vabishchevich2014a}.

In this work, using Samarskii's regularization principle for operator-difference sche\-mes we construct splitting schemes with respect to physical processes for doub\-le-porosity poroelasticity problems. 
The paper is organized as follows. After a brief description of the mathematical problem in Section 2, we introduce the variational formulation of problem in Section 3. Here, we obtain a priori estimate that ensures the stability of the solution of problem. Discretization in space and time are performed in Sections 4 and 5, respectively.    Splitting schemes are constructed in Sections 6 and 7. Numerical experiments that test the convergence and stability of considered schemes are shown in Section 8. Conclusions follow. 

\section{Mathematical Model}
\label{sec:model}
 
We consider double-porosity poroelasticity problems arising when a fluid flows through homogeneous and isotropic fractured medium where fractures and porous matrix represent two overlapping continua \cite{Bai1993,Barenblatt1960}. In these continua two fluid flow processes with two different pressure fields occur. Therefore, the poroelasticity equations can be expressed as
\begin{equation}
\label{eq:u}
\div \bm \sigma (\bm u) - \alpha_1 \grad p_1 - \alpha_2 \grad p_2 = 0,
\end{equation}
\begin{equation}
\label{eq:p1}
\begin{split}
\beta_1 \pdt{p} + \alpha_1 \pdt{\div \bm u} -& \div \left(\frac{\bm k_1}{\eta} \grad p_1 \right) + \gamma (p_1 - p_2) = f_1(\bm x,t),
\end{split}
\end{equation}
\begin{equation}
\label{eq:p2}
\begin{split}
\beta_2 \pdt{p} + \alpha_2 \pdt{\div \bm u} -& \div \left(\frac{\bm k_2}{\eta} \grad p_2 \right) + \gamma (p_2 - p_1) = f_2(\bm x,t).
\end{split}
\end{equation}
Here, $\bm u$ is the displacement vector, $p_1$ is the fluid pressure in pores, and $p_2$ is fluid pressure in fractures. The stress tensor $\bm \sigma$ is defined by the expression
\[
\bm \sigma = 2 \mu \bm \varepsilon (\bm u) + \lambda \div \bm u \bm I, 
\]
where $\mu$ is the shear modulus, $\lambda$ are the Lame coefficient, $\bm I$ is the unit tensor, and $\bm \varepsilon$ is the strain tensor:
\[
\bm \varepsilon  = \frac{1}{2}(\grad \bm u + \grad \bm u^T).
\]
The other notation is as follows: $\alpha$ is Biot coefficients, $\beta = 1/M$, $M$ is the Biot modulus, $\bm k_1$,  $\bm k_2$ are the permeability tensors, $\eta$ is the viscosity of the fluid, $\gamma$ is the exchange parameter, and $f_1$, $f_2$ are the functions describing given fluid sources (sinks). The subscripts $1$ and $2$ represent notation associated with pores and fractures, respectively. 

The system (\ref{eq:u})--(\ref{eq:p2}) is considered in a bounded domain $\Omega$ with a boundary $\Gamma$, on which, for simplicity, we set the following homogeneous conditions for displacements 
\begin{equation}
\label{eq:bc_u}
\quad \bm u = 0, \quad \bm x \in \Gamma_D,  \quad  \bm \sigma \bm n = 0, \quad \bm x \in \Gamma_N .
\end{equation}
For pressures we set
\begin{equation}
\label{eq:bc_p1}
p_1 = 0, \quad \bm x \in \Gamma_N, \quad -\frac{\bm k_1}{\eta} \pd{ p_1}{ n} = 0, \quad \bm x \in \Gamma_D,
\end{equation}
\begin{equation}
\label{eq:bc_p2}
p_2 = 0, \quad \bm x \in \Gamma_N, \quad -\frac{\bm k_2}{\eta} \pd{p_2}{n} = 0, \quad \bm x \in \Gamma_D.
\end{equation}
Here, $\bm n$ is the unit normal to the boundary, $\Gamma = \Gamma_D + \Gamma_N$.
In addition, we specify the initial conditions for pressures as
\begin{equation}
\label{eq:ic_p}
p_1(\bm x, 0) = s_1(\bm x), \quad p_2(\bm x, 0) = s_2(\bm x), \quad \bm x \in \Omega.
\end{equation}
The initial-boundary problem (\ref{eq:u})--(\ref{eq:ic_p}) for coupled para\-bolic and elliptic equations is the basis for considering fluid flow in deformable  fractured porous medium.   

\section{Variational Formulation}

For the numerical solution, we use finite element approximation in space \cite{brenner2008mathematical,KnabnerAngermann2003}, so we need obtain a variational formulation of  problem (\ref{eq:u})--(\ref{eq:ic_p}). For scalar quantities, let us introduce the Hilbert space $L_2(\Omega)$ with the scalar product and norm given as
\[
(u,v) = \int_{\Omega} u(\bm x) v(\bm x), \quad \| u \| = (u,u)^{1/2}.
\]
For vector quantities, we use $\bm L_2(\Omega) = [L_2(\Omega)]^m$, where $m=2,3$ is the dimension of domain $\Omega$. Let $H^1(\Omega)$ and $\bm H^1(\Omega)$ be the Sobolev spaces.  Next, we define the subspaces of scalar and vector functions
\[
Q = \lbrace q \in H^1(\Omega): \, q(\bm x) = 0, \, \bm x \in \Gamma_D \rbrace,
\]
\[
\bm V = \lbrace \bm v \in \bm H^1(\Omega): \, \bm v(\bm x) = 0, \, \bm x \in \Gamma_D \rbrace.
\]

After multiplying (\ref{eq:u}), (\ref{eq:p1}), and (\ref{eq:p2}) by test functions $\bm v \in \bm V$  and $q_1, q_2 \in Q$, respectively, and integrating by parts to eliminate the second order derivatives, we come to the following variational problem:
Find $\bm u\in \bm V$, $p_1, \, p_2 \in Q$ such that
\begin{equation}
\label{eq:u_var}
a(\bm u, \bm v) + \alpha_1 \, g(p_1,\bm v) + \alpha_2 \, g(p_2, \bm v) = 0,  \quad  
\forall \bm  v \in \bm V,
\end{equation}
\begin{equation}
\label{eq:p1_var}
\begin{split}
c_1 \left(\pdt{p_1},q_1\right) & + \alpha_1 \, d\left(\pdt{\bm u},q_1\right) + b_1(p_1,q_1)  + \gamma (p_1 - p_2 , q_1) 
 = (f_1, q_1), \quad \forall  q_1 \in Q,
\end{split}
\end{equation}
\begin{equation}
\label{eq:p2_var}
\begin{split}
c_2 \left(\pdt{p_2},q_2\right) & + \alpha_2 \, d\left(\pdt{\bm u},q_2\right) + b_2(p_2,q_2) + \gamma (p_2 - p_1 , q_2) 
 = (f_2, q_2), \quad  \forall q_2 \in Q.
\end{split}
\end{equation}
Here, the bilinear forms are defined as
 \[
 a(\bm u, \bm v) = \int_{\Omega} \bm \sigma (\bm u) \bm \varepsilon (\bm v)\, d\bm x, 
 \]
 \[
 g(p,\bm v) = \int_{\Omega} \grad p \, \bm v \, d\bm x, \, \quad 
 d(\bm u, q) = \int_{\Omega} \div \bm u \, q \, d\bm x,
 \]
 \[
c_1(p,q) =   \beta_1 \int_{\Omega} p \, q \, d\bm x, \quad 
c_2(p,q) =   \beta_2 \int_{\Omega} p \, q \, d\bm x,
\]
\[
b_1(p,q) =   \int_{\Omega} \frac{\bm k_1}{\eta} \grad p \, \grad q \, d\bm x, 
\]
\[
b_2(p,q) =   \int_{\Omega} \frac{\bm k_2}{\eta} \grad p \, \grad q \, d\bm x.
\]
The initial conditions (\ref{eq:ic_p}) are set as:
\begin{equation}
\label{eq:ic_var}
\begin{split}
(p_1(0),q_1) = (s_1,q_1), \quad (p_2(0),q_2) = (s_2,q_2), \quad q_1, q_2 \in Q.
\end{split}
\end{equation}

The bilinear form  $a(\cdot, \cdot)$ is symmetric and positive definite:
\[
a(\bm u,\bm v) = a(\bm v,\bm u), \, a(\bm u, \bm u) \geq \delta_a \| \bm u \|^2, \, \delta_a > 0, \, \bm u, \bm v \in \bm V. 
\]
Taking into account this, the form  is associated with a Hilbert space $H_a$ with the following inner product and norm:
\[
(\bm u, \bm v)_a = a(\bm v, \bm u), \quad \| \bm u \|_a =  (a(\bm u,\bm u))^{1/2}. 
\]
The norm of the adjoint of the space $H_a$ is denoted by $\| \bm u \|_{*,a}$. Note, that forms $b_1(\cdot, \cdot)$, $b_2(\cdot, \cdot)$, $c_1(\cdot, \cdot)$, and $c_2(\cdot, \cdot)$ are also symmetric and positive definite.

According to the Green's formula, we have
\[
\int_{\Omega} \div \bm u \, p \, d\bm x = - \int_{\Omega} \grad p \, \bm u \, d\bm x + \int_{\Gamma}  \bm u \bm n \, p \, d \bm x.
\] 
Then, in view of the boundary conditions (\ref{eq:bc_u})--(\ref{eq:bc_p2}), forms $g(\cdot, \cdot)$ and $d(\cdot, \cdot)$ are related as 
\[
d(\bm u,p) = - \, g(p,\bm u),  \quad p \in Q, \quad \bm u \in \bm V.
\]

Now, we derive the simplest a priori estimates for the solution of problem  (\ref{eq:u_var})--(\ref{eq:ic_var}). 
Setting $\bm v  = \partial \bm u/ \partial t$ in (\ref{eq:u_var}), $q_1 = p_1$ in (\ref{eq:p1_var}), and $q_2 = p_2$ in (\ref{eq:p2_var}), summing up these equations and taking into account that, for example,
\[
a\left(\bm u, \pdt{\bm u}\right) = \frac{1}{2} \dt{} a(\bm u, \bm u),
\]
 we obtain 
\[
\begin{split}
\frac{1}{2} \dt{} \big ( a (\bm u, \bm u) & + c_1( p_1, p_1) + c_2( p_2, p_2 ) \big )   + b_1(p_1,p_1) + b_2(p_2, p_2)  \\ 
& + \gamma \| p_1 - p_2 \|^2 = (f_1,p_1) + (f_2, q_2).
\end{split}
\]
For the right-hand side we use the following inequalities
\[
( f_l, p_l) \leq \| p_l \|_{b_l}^2  + \frac{1}{4}  \| f_l  \|_{*,b_l}^2, \quad l = 1,2.
\]
Thus, we have
\[
\dt{} ( \| \bm u \|_a^2 +  \| p_1 \|_{c_1}^2 +   \| p_2 \|_{c_2}^2)  \leq \frac{1}{2} \left(  \| f_1 \|_{*,b_1}^2+  \| f_1 \|_{*,b_1}^2 \right).
\]

Using the initial condition (\ref{eq:ic_var}), we compute the initial displacement $\bm u_0(\bm x)$ 
\[
a(\bm u_0, \bm v) + \alpha_1 \, g(s_1,\bm v) + \alpha_2 \, g(s_2, \bm v) = 0,  \quad   v \in V.
\]
Integration with respect to time gives the estimate:
\begin{equation}
\label{eq:apriori_var}
\begin{split}
| \bm u \|_a^2 +   \| p_1 \|_{c_1}^2  & +   \| p_2 \|_{c_2}^2  \leq \| \bm u_0 \|_a^2 +   \| s_1 \|_{c_1}^2   \\ & + \| s_2 \|_{c_2}^2   + \frac{1}{2} \int_0^t \left(  \| f_1(\xi) \|_{*,b_1}^2+  \| f_2(\xi) \|_{*,b_2}^2 \right) d\xi.
\end{split}
\end{equation}
using the notation $f(\cdot,t) = f(t)$.

A priori estimate  (\ref{eq:apriori_var})  ensures the stability of the solution of problem (\ref{eq:u_var})-- (\ref{eq:ic_var}) with respect to the initial data and the right-hand side. 
Similar estimates can be obtained for the solution of discrete problem.

\section{Finite Element Approximation}

Now, we approximate our problem in space using finite element methods.
First, we construct a computational mesh $\Omega_h = \lbrace \omega_1, \omega_2, \dots, \omega_N \rbrace $ of domain $\Omega$. Here, $N$ is the number of cells $\omega$,  $h = \max_{\omega \in \Omega_h} h_{\omega}$, where $h_{\omega}$ is the diameter of  circle inscribed in a cell $\omega$. 

Then, in the mesh we define  spaces of conformal finite elements for scalar and vector functions $Q_h \subset Q$ and $\bm V_h \subset \bm V$ and restrict the variational problem (\ref{eq:u_var})--(\ref{eq:ic_var}) to these spaces: find $\bm u_h \in \bm V_h$ and $p_{1,h}, p_{2,h} \in Q_h$ such that
\begin{equation}
\label{eq:u_discr}
a(\bm u_h, \bm v) + \alpha_1 g(p_{1,h},\bm v) + \alpha_2  g(p_{2,h}, \bm v) = 0,  \, \forall  \bm  v \in \bm V_h,
\end{equation}
\begin{equation}
\label{eq:p1_discr}
\begin{split}
c_1 \left(\pdt{p_{1,h}},q_1\right) & + \alpha_1 d\left(\pdt{\bm u_h},q_1\right) +  b_1(p_{1,h},q_1) \\ 
& + \gamma (p_{1,h} - p_{2,h} , q_1) = (f_1, q_1), \, q_1 \in Q_h,
\end{split}
\end{equation}
\begin{equation}
\label{eq:p2_discr}
\begin{split}
c_2 \left(\pdt{p_{2,h}},q_2\right) & + \alpha_2 d\left(\pdt{\bm u_h},q_2\right) + b_2(p_{2,h},q_2) \\ 
& + \gamma (p_{2,h} - p_{1,h} , q_2)= (f_2, q_2), \, q_2 \in Q_h,
\end{split}
\end{equation}
\begin{equation}
\label{eq:ic_discr}
\begin{split}
(p_{1,h}(0),q_1) = (s_1,q_1), \quad (p_{2,h}(0),q_2) = (s_2,q_2),   \,  q_1, q_2 \in Q_h.
\end{split}
\end{equation}

For further consideration, it is convenient to use the operator formulation of problem (\ref{eq:u_discr})-- (\ref{eq:ic_discr}). We define finite-dimensional operators $A$, $B_1$, $B_2$, $C_1$, $C_2$, $D$, $G$ related to the corresponding bilinear forms by setting, for example,
\[
(A \bm u_h, \bm v) = a(\bm u_h, \bm v), \quad \forall \bm u_h, \bm v \in \bm V_h.
\]
As a result, we come from problem (\ref{eq:u_discr})--(\ref{eq:ic_discr}) to the following Cauchy problem for a system of equations: 
\begin{equation}
\label{eq:u_oper}
A \bm u_h + \alpha_1 G p_{1,h} + \alpha_2 G p_{2,h} = 0,  
\end{equation}
\begin{equation}
\label{eq:p1_oper}
\begin{split}
C_1 \dt{p_{1,h}}  + \alpha_1 D \dt{\bm u_h} +& B_1 p_{1,h} + \gamma E (p_{1,h} - p_{2,h}) = f_{1,h},
\end{split}
\end{equation}
\begin{equation}
\label{eq:p2_oper}
\begin{split}
C_2 \dt{p_{2,h}}  + \alpha_2 D \dt{\bm u_h} +& B_2 p_{2,h} + \gamma E (p_{2,h} - p_{1,h}) = f_{2,h},
\end{split}
\end{equation}
\begin{equation}
\label{eq:ic_oper}
p_{1,h}(0) = s_{1,h}, \quad p_{2,h}(0)  = s_{2,h}.
\end{equation}
where $E$ is the identity operator and
\[
(f_{l,h},q_l) = (f_l,q_l), \, (s_{l,h},q_1) = (s_l,q),  \, \forall q_l \in Q_h, \, l = 1,2.
\]

The operator $A$, $B_1$, $B_2$, $C_1$ and $C_2$ are self-adjoint and positive definite, for example,
\[
A = A^* > 0,
\]
while $D$ and $G$ are adjoint with to each other with an opposite sign:
\[
D = - G^*.
\]

Note that for the operator formulation (\ref{eq:u_oper})--(\ref{eq:ic_oper}) we can derive a discrete analogue of estimate (\ref{eq:apriori_var}):
\begin{equation}
\label{eq:apriori_oper}
\begin{split}
\| \bm u_h \|_A^2 +&   \| p_{1,h} \|_{C_1}^2 +   \| p_{2,h} \|_{C_2}^2 \leq \| \bm u_{0,h} \|_A^2 + \| s_{1,h} \|_{C_1}^2 +   \| s_{2,h} \|_{C_2}^2   \\ +& \frac{1}{2} \int_0^t \left(  \| f_{1,h}(\xi) \|_{*,B_1}^2+  \| f_{2,h}(\xi) \|_{*,B_2}^2 \right) d\xi.
\end{split}
\end{equation}

Now, introducing a vector of pressures: $\bm p_h = \lbrace p_{1,h}$, $p_{2,h} \rbrace$, from (\ref{eq:u_oper})--(\ref{eq:ic_oper})  we obtain
\begin{equation}
\label{eq:u_vec}
A \bm u_h + \bm G \bm p_h = 0,  
\end{equation}
\begin{equation}
\label{eq:p_vec}
\bm C \dt{\bm p_h}  + \bm D \dt{\bm u_h}+ \bm B \bm p_h = \bm f_h,
\end{equation}
\begin{equation}
\label{eq:ic_vec}
\bm p_h(0) = \bm s_h.
\end{equation}
Here
\begin{equation}
\label{eq:vec_opers}
\begin{split}
\bm G &= \begin{pmatrix}
\alpha_1 G & \alpha_2 G \\ 
\end{pmatrix}, \quad
\bm D = \begin{pmatrix}
\alpha_1 D \\ 
\alpha_2 D
\end{pmatrix}, \\
\bm C & = \begin{pmatrix}
C_1 & 0\\ 
0 & C_2
\end{pmatrix}, \quad
\bm B = \begin{pmatrix}
B_1 + \gamma E  &  - \gamma E  \\ 
-  \gamma E  & B_2 +  \gamma E  
\end{pmatrix}, 
\end{split}
\end{equation}
and 
\[
\bm f_h = \lbrace f_{1,h}, f_{2,h} \rbrace, \quad \bm s_h = \lbrace s_{1,h}, s_{2,h} \rbrace.
\]
The operators $\bm C$ and $\bm B$ are self-adjoint and positive definite, and $\bm G^* = - \bm D$ in the sense of the equality $(\bm G \bm p_h, \bm u_h) = - (\bm p_h, \bm D \bm u_h)$.
Then, the solution of problem (\ref{eq:u_vec})--(\ref{eq:ic_vec}) satisfies the a priori estimates
\[
\| \bm u_h \|_A^2 +   \| \bm p_h \|_{\bm C}^2 \leq \| \bm u_{0,h} \|_A^2 + \| \bm s_h \|_{\bm C}^2 + 
\frac{1}{2} \int_0^t   \|\bm f_h(\xi) \|_{*,\bm B}^2d\xi.
\]
Note that this estimate is similar to the apriori estimates (\ref{eq:apriori_var}), (\ref{eq:apriori_oper}).

\section{Time Discretisation }

For the discretization in time we use a uniform grid with a step $\tau > 0$.
Let  $\bm u^n = \bm u_h(\bm x, t^n)$, $\bm p^n = p_h(\bm x, t^n)$, where $t^n = n \tau$, $n=0,1, \dots$. 
To obtain an approximate solution of problem (\ref{eq:u_oper})--(\ref{eq:ic_oper}) we use the standard two-level scheme with weights:
\begin{equation}
\label{eq:u_coupled}
A \bm u^{n+1} + \bm G \bm p^{n+1} = 0,  
\end{equation}
\begin{equation}
\label{eq:p_coupled}
\begin{split}
\bm C \frac{\bm p^{n+1}-p^n}{\tau} + \bm D \frac{\bm u^{n+1}-\bm u^n}{\tau} + \bm B \bm p_{\theta}^{n+1}  = \bm f_{\theta}^{n+1}, 
\end{split}
\end{equation}
where
\[
\bm p_{\theta}^{n+1} = \theta
 \bm p^{n+1} + (1-\theta) \bm p^n, \quad \bm f_{\theta}^{n+1} = \bm f_h(t_{\theta}^{n+1})
\]
and  $0 \leq \theta \leq 1$.
The initial condition is written as
\begin{equation}
\label{eq:ic_coupled}
(\bm p^0, \bm q) = (\bm s,\bm q).
\end{equation}

\begin{thm}\label{t-1}
For $\theta \geq 0.5$ the solution of problem (\ref{eq:u_coupled})--(\ref{eq:ic_coupled}) satisfies the a priori estimate
\begin{equation}\label{eq:apriori_coupled}
\begin{split}
 \|\bm u^{n+1}\|_A^2 + \|\bm p^{n+1}\|_{\bm C}^2 \leq  
 \|\bm u^{n}\|_A^2 +& \|\bm p^{n}\|_{\bm C}^2 + \frac{\tau}{2} \|\bm f_{\theta}^{n+1} \|_{\bm B^{-1}}^2 .
 \end{split}
\end{equation} 
\end{thm}
\begin{pf}
In view if linearity, multiplying (\ref{eq:u_coupled}) by ${\displaystyle \frac{\bm u^{n+1}-\bm  u^n}{\tau}}$ and (\ref{eq:p_coupled}) by $p_{\theta}^{n+1}$, we get
\[
\left( A \bm u_{\theta}^{n+1}, \frac{\bm u^{n+1} - \bm u^n}{\tau}\right) + 
\left( \bm G \bm p_{\theta}^{n+1}, \frac{\bm u^{n+1} - \bm u^n}{\tau}\right) = 0,
\]
\[
\begin{split}
 \left(\bm C \frac{\bm p^{n+1} - \bm p^n}{\tau},  \bm p_{\theta}^{n+1} \right) 
 + \left( \bm D \frac{\bm u^{n+1} - \bm u^n}{\tau},  \bm p_{\theta}^{n+1} \right)
 + (\bm B \bm p_{\theta}^{n+1} ,  \bm p_{\theta}^{n+1}) = (\bm f_{\theta}^{n+1} ,  \bm p_{\theta}^{n+1}),
\end{split}
\]
where 
\[
 \bm u_{\theta}^{n+1} = \theta \bm u^{n+1}+(1-\theta) \bm u^n.
\]
Summing up  these equations, we obtain
\[
\begin{split}
 \left(A \bm u_{\theta}^{n+1}, \frac{\bm u^{n+1} - \bm u^n}{\tau} \right)
 + \left(\bm C \frac{\bm p^{n+1} - \bm p^n}{\tau},  \bm p_{\theta}^{n+1} \right)  
 + (\bm B \bm p_{\theta}^{n+1} ,  \bm p_{\theta}^{n+1}) = (\bm f_{\theta}^{n+1} ,  \bm p_{\theta}^{n+1}).
 \end{split}
\]
Taking into account inequality
\[
 (\bm f_{\theta}^{n+1}, \bm p_{\theta}^{n+1}) \leq \|\bm p_{\theta}^{n+1}\|_{\bm B}^2 + \frac{1}{4} \|\bm f_{\theta}^{n+1}\|_{\bm B^{-1}}^2 ,
\]
gives
\[
\begin{split}
 \left(A \bm u_{\theta}^{n+1}, \frac{\bm u^{n+1} - \bm u^n}{\tau} \right)
+ \left(\bm C \frac{\bm p^{n+1} - \bm p^n}{\tau},  \bm p_{\theta}^{n+1} \right) 
\leq \frac{1}{4} ||\bm f_{\theta}^{n+1}||_{\bm B^{-1}}^2 .
\end{split}
\]
Using the equality
\[
\begin{split}
\bm v^{n+1}_{\theta} = \theta \bm v^{n+1} + (1-\theta) \bm v^n = 
 \frac{\bm v^{n+1} +  \bm v^n}{2} + \left(\theta -\frac{1}{2}\right) (\bm v^{n+1} - \bm v^n) 
 \end{split}
\]
and taking into account that any self-adjoint operator $A$  satisfies 
\[
 (A(\bm u+ \bm v), \bm u- \bm v) = (A \bm u, \bm u) - (A \bm v, \bm v),
\]
we obtain
\[
\begin{split}
 \frac{1}{2} \left( || \bm u^{n+1}||^2_A - ||\bm u^n||^2_A \right)  & + 
 \frac{1}{2} \left( ||\bm p^{n+1}||^2_{\bm C} - ||\bm p^n||^2_{\bm C} \right) \\
  & + \left(\theta -\frac{1}{2}\right) \left( ||\bm u^{n+1} - \bm u^n||^2_A  + ||\bm p^{n+1} - \bm p^n||^2_{\bm C} \right)
  \leq \frac{\tau}{4} ||\bm f_{\theta}^{n+1}||_{\bm B^{-1}}^2 .
\end{split}
\]
For $\theta \geq 0.5$, this yields the required estimate (\ref{eq:apriori_coupled}).
\end{pf}

Note that in the numerical implementation of weighted scheme (\ref{eq:u_coupled})--(\ref{eq:ic_coupled}) $\bm u^{n+1}$, $p_1^{n+1}$, and $p_2^{n+1}$ are determined at each time level by solving the coupled system:
\[
A \bm u^{n+1} + \alpha_1 G p_1^{n+1} + \alpha_2 G p_2^{n+1} = 0,
\]
\[
\begin{split}
C_1 p_1^{n+1} + \alpha_1 D \bm u^{n+1} + \theta \tau (B_1 + \gamma E) p_1^{n+1}  - \theta \tau \gamma E p_2^{n+1}  =  \psi_1^{n+1}, 
\end{split}
\]
\[
\begin{split}
C_2 p_2^{n+1} + \alpha_2 D \bm u^{n+1} + \theta \tau (B_2 + \gamma E) p_2^{n+1}  - \theta \tau \gamma E p_1^{n+1}  =  \psi_2^{n+1}, 
\end{split}
\]
with the corresponding $\psi_1^{n+1}$ and $\psi_2^{n+1}$.  Various special numerical algorithms can be used to solve this system \cite{Alexsson2013,Gaspar2007a}. 

Another opportunity is to construct splitting schemes:
\[
A \bm u^{n+1} = \varphi_{\bm u}^{n+1},
\]
\[
C_1 p_1^{n+1} + \theta \tau B_1 p_1^{n+1}  =  \varphi_{p_1}^{n+1}, 
\]
\[
C_2 p_2^{n+1} + \theta \tau B_2 p_2^{n+1}  =  \varphi_{p_2}^{n+1}, 
\]
where the transition to a new time level involves the solution of separate equations for displacements and pressures.

\section{Incomplete Splitting Scheme}

To construct splitting schemes, we express $\bm u = \bm u_h$ from (\ref{eq:u_vec}) and substitute the result into (\ref{eq:p_vec}), which leads to a single equation for $\bm p = \bm p_h$
\[
- \bm D A^{-1} \bm G \dt{\bm p} + \bm C \dt{\bm p} + \bm B \bm p = \bm f.
\]
This equation can be written as
\begin{equation}
\label{eq:p_single}
 \widetilde{\bm B} \dt{\bm p} + \widetilde{\bm A}  \bm p = \bm f, 
\end{equation} 
where $ \widetilde{\bm A} = \bm B$, and the operator $\widetilde{\bm B}$ is equal to sum of two self-adjoint, positive definite operators:
\begin{equation}
\label{eq:B}
 \widetilde{\bm B} = \widetilde{\bm B}_0 +  \widetilde{\bm B}_1, \quad \widetilde{\bm B}_0 = \bm C,
 \quad \widetilde{\bm B}_1 = - \bm D A^{-1} \bm G.
\end{equation}
The operators $\widetilde{\bm A}, \widetilde{ \bm B}$ are  self-adjoint and positive definite. 

Under the following constraint
\begin{equation}\label{eq:constraint}
 \widetilde{\bm B}_1 \leq \gamma_{B} \widetilde{\bm B}_0, \quad \gamma_B > 0, 
\end{equation} 
for the numerical solution of problem (\ref{eq:ic_vec}), (\ref{eq:p_single}), we can use a thee-level explicit-implicit scheme with weights \cite{Gaspar2014}:
\begin{equation}\label{eq:ode_exim}
\begin{split}
 \widetilde{\bm B}_0 \Bigl( \theta \frac{\bm p^{n+1} - \bm p^{n}}{\tau }  + 
 (1-\theta) \frac{\bm p^{n} - \bm p^{n-1}}{\tau } \Bigr)  +
 \widetilde{\bm B}_1 \frac{\bm p^{n} - \bm p^{n-1}}{\tau }  
  + \widetilde{\bm A} \bm p^{n+1} =\bm f^{n+1}.
\end{split}
\end{equation} 
To calculate the first step, we can apply the two-level scheme
\begin{equation}\label{eq:ic_exim}
\bm p^0 = \bm p_0,
 \quad \widetilde{\bm B} \frac{\bm p^1-\bm p^0}{\tau} + \widetilde{\bm A} \bm p^1 = \bm f^1.
\end{equation} 
The value of   $\theta$ is determined by the stability conditions for difference scheme (\ref{eq:ode_exim}), (\ref{eq:ic_exim}).

Taking into account equalities
\[
 \frac{\bm p^{n+1} - \bm p^{n}}{\tau} =  \frac{\bm p^{n+1} - \bm p^{n-1}}{2\tau} + \frac{\tau }{2} \frac{\bm p^{n+1} -2 \bm p^{n} + \bm p^{n-1}}{\tau^2} ,
\] 
\[
 \frac{\bm p^{n} - \bm p^{n-1}}{\tau} =  \frac{\bm p^{n+1} - \bm p^{n-1}}{2\tau} - \frac{\tau }{2} \frac{\bm p^{n+1} -2 \bm p^{n} + \bm p^{n-1}}{\tau^2} ,
\]
\[
\bm p^{n+1} = \tau \frac{\bm p^{n+1}-\bm p^n}{\tau} + \bm p^n,
\]
we can write the scheme (\ref{eq:ode_exim}) in the canonical form of three-level operator-difference schemes
\begin{equation}
\label{eq:three-level}
\begin{split}
\widetilde{\bm C} \frac{\bm p^{n+1} - \bm p^{n-1}}{2\tau} + \widetilde{\bm D} \frac{\bm p^{n+1} -2 \bm p^{n} + \bm p^{n-1}}{\tau^2}  + \widetilde{\bm A} \bm p^n = \bm f^n.
\end{split}
\end{equation}
where $\widetilde{\bm C}$, $\widetilde{\bm D}$ are the self-adjoint and positive definite operators given by
\begin{equation}
\label{eq:D}
\widetilde{\bm C} = \widetilde{\bm B} + \tau \widetilde{\bm A}, \quad  \widetilde{\bm D} = \frac{\tau}{2} \left( (2 \theta -1) \widetilde{\bm B}_0 - \widetilde{\bm B}_1 \right) + \frac{\tau^2}{2} \widetilde{\bm A}.
\end{equation}

Our subsequent analysis is based on the following general statement from the stability (well-posedness) theory of three-level operator-difference schemes \cite{Samarskii1989,Samarskii2002}. 

\begin{lm}\label{lem:1}
For 
\begin{equation}
\label{eq:three-level-condition}
\widetilde{\bm D} > \frac{\tau^2}{4}\widetilde{\bm A},
\end{equation}
scheme (\ref{eq:three-level}) is unconditionally stable and its solution satisfies the estimate
\[
  \mathcal{E}^{n+1} \leq 
  \mathcal{E}^{n} +
  \frac{\tau}{2} \left \| \bm f^{n+1} \right \|^2_{\widetilde{ \bm C}^{-1}},
\]
where
\[
  \mathcal{E}^{n} = 
  \left \| \frac{\bm p^{n} + \bm p^{n-1}}{2} \right \|^2_{\widetilde{\bm A}} +
  \left \| \frac{\bm p^{n} - \bm p^{n-1}}{\tau} \right \|^2_{\widetilde{\bm D}-\frac{\tau^2}{4} \widetilde{\bm A}} .
\]
\end{lm}

Then, stability condition (\ref{eq:three-level-condition}) holds if
\[
(2 \theta - 1) \widetilde{\bm B}_0 - \widetilde{\bm B}_1 \geq 0 .
\]
In view of inequalities (\ref{eq:constraint}), this is achieved if we choose
\begin{equation}
\label{eq:theta_b}
2\theta \geq 1 + \delta,
\end{equation}
where $\delta$ is determined from (\ref{eq:constraint}) as the maximum eigenvalue of spectral problem $\delta = \nu_{\max}$:
\[
 \widetilde{ \bm B}_1 \bm p = \nu  \widetilde{ \bm B}_0 \bm p .
\]
Taking into account (\ref{eq:B}) yields
\[
 - \bm D A^{-1} \bm G \bm p = \nu  \bm C \bm p .
\] 
We set $\bm u = - A^{-1} \bm G \bm p$ and come to the problem 
\[
A \bm u + \bm G \bm p =0, \quad \bm D \bm u = \nu  \bm C \bm p.
\]
Excluding $\bm p$, we have 
\[
 \bm G \bm C^{-1}  \bm D \bm u + \nu  A \bm u = 0 .
\]
Taking into account (\ref{eq:vec_opers}), we get the Cosserat spectrum problem
\cite{Chigonkov,mikhlin1973spectrum} 
\begin{equation}
\label{eq:eigen_b}
(\alpha_1^2 + \alpha_2^2) G \bm C^{-1} D \bm u + \nu  A \bm u = 0.
\end{equation}
In principle, the maximum eigenvalue in (\ref{eq:eigen_b})  depends on the physical parameters of the differential problem ($\alpha_i, \beta_i, \ i = 1,2$, $\mu, \lambda$) and the domain, but not on the time steps.

Combining (\ref{eq:B}) and (\ref{eq:ode_exim}), we obtain the following incomplete splitting scheme:
\begin{equation}
\label{eq:u_split_incomplete}
A \bm u^{n+1} + \bm G \bm p^{n} = 0,  
\end{equation}
\begin{equation}
\label{eq:p_split_incomplete}
\begin{split}
\bm C \Big ( \theta \frac{\bm p^{n+1}-\bm p^n}{\tau}  + (1-\theta)\frac{\bm p^{n}-\bm p^{n-1}}{\tau} \Big ) + \bm D \frac{\bm u^{n+1}-\bm u^n}{\tau} + \bm B \bm p^{n+1}  = \bm f^{n+1}, 
\end{split}
\end{equation}
where the solution is determined by solving separately  the elasticity problem (\ref{eq:u_split_incomplete}) and the double porosity problem (\ref{eq:p_split_incomplete}).
The stability of this splitting scheme can be described by the following theorem:
\begin{thm}
 Scheme (\ref{eq:u_split_incomplete}), (\ref{eq:p_split_incomplete}) is unconditionally stable for  $2\theta \geq 1 + \delta$,
where $\delta = \nu_{\max}$ is the maximum eigenvalue of spectral problem  (\ref{eq:eigen_b}).
\end{thm}  

\section{Full Splitting Scheme}

In the splitting scheme with respect to physical processes (\ref{eq:u_split_incomplete}), (\ref{eq:p_split_incomplete}) 
on each time level we solve two separate problems: the displacement problem (\ref{eq:u_split_incomplete}) and the coupled problem for pressures in pores and fractures (\ref{eq:p_split_incomplete}).
It makes sense to build a splitting scheme, where we have separate problems for pressures as well.

We can split the scheme (\ref{eq:p_single}) further separating the diagonal part of operator $\widetilde{\bm A}$ as follows
\begin{equation}
\label{eq:A}
 \widetilde{\bm A} = \widetilde{\bm A}_0 +  \widetilde{\bm A}_1,
\end{equation}
where
\[
 \widetilde{\bm A}_0 = \begin{pmatrix}
B_1 + \gamma E  &  0  \\ 
0  & B_2 +  \gamma E  
\end{pmatrix},  \quad 
\widetilde{\bm A}_1 = \begin{pmatrix}
0 &  - \gamma E  \\ 
-  \gamma E  & 0 
\end{pmatrix}. 
\] 

Under the assumption
\begin{equation}
\label{eq:constraint_A}
\widetilde{\bm A}_0 - \widetilde{\bm A}_1 \geq 0
\end{equation}
which follows from positive-definiteness of the operator $\widetilde{\bm A}$,
we can use the following explicit-implicit scheme
\begin{equation}\label{eq:ode_exim_full}
\begin{split}
 &\widetilde{\bm B}_0 \Big ( \theta \frac{\bm p^{n+1} - \bm p^{n}}{\tau }  + 
 (1-\theta) \frac{\bm p^{n} - \bm p^{n-1}}{\tau } \Big )  \\ 
 &+\widetilde{\bm B}_1 \frac{\bm p^{n} - \bm p^{n-1}}{\tau }  
  + \widetilde{\bm A}_0 \bm p^{n+1} +  \widetilde{\bm A}_1 \bm p^{n} =\bm f^{n+1}.
\end{split}
\end{equation}

Similar to (\ref{eq:ode_exim}) scheme (\ref{eq:ode_exim_full}) can be written in the canonical form of three-level operator difference schemes (\ref{eq:three-level}) with
\[
\widetilde{\bm C} = \widetilde{\bm B} + \tau \widetilde{\bm A}_0, \quad
 \widetilde{\bm D} = \frac{\tau}{2} \left( (2 \theta -1) \widetilde{\bm B}_0 - \widetilde{\bm B}_1 \right) + \frac{\tau^2}{2} \widetilde{\bm A}_0,
\]
and the analysis of the resulting scheme can be based on Lemma \ref{lem:1}.
In this case,  the stability condition (\ref{eq:three-level-condition}) takes the form
\[
2 \left( (2 \theta -1) \widetilde{\bm B}_0 - \widetilde{\bm B}_1 \right) + \tau ( 2 \widetilde{\bm A}_0 -  \widetilde{\bm A}) > 0.
\]
Taking into account (\ref{eq:A}), (\ref{eq:constraint_A}), the stability condition becomes $2 \theta \geq 1 + \delta$, where $\delta$ is determined from (\ref{eq:eigen_b}).

Finally, combining (\ref{eq:B}), (\ref{eq:A}), and (\ref{eq:ode_exim_full}), we have the full splitting scheme:
\begin{equation}
\label{eq:u_split}
 A \bm u^{n+1}  +  \alpha_1 G p_1^{n} + \alpha_2 G p_2^n = 0, 
\end{equation} 
\begin{equation}
\label{eq:p1_split}
\begin{split}
\alpha_1 D \frac{\bm u^{n+1} - \bm u^{n}}{\tau}  & + B_1 p_1^{n+1} + \gamma( p_1^{n+1} - p_2^n)  \\ 
& +  C_1 \Bigl (\theta \frac{p_1^{n+1} - p_1^{n}}{\tau} + 
 (1-\theta)\frac{p_1^{n} - p_1^{n-1}}{\tau} \Bigr)  = f_1^{n+1},
\end{split}
\end{equation} 
\begin{equation}
\label{eq:p2_split}
\begin{split}
\alpha_2 D \frac{\bm u^{n+1} - \bm u^{n}}{\tau}  & + B_2 p_2^{n+1} + \gamma( p_2^{n+1} - p_1^n)  \\ 
& +  C_2 \Bigl (\theta \frac{p_2^{n+1} - p_2^{n}}{\tau} + 
 (1-\theta)\frac{p_2^{n} - p_2^{n-1}}{\tau} \Bigr)  = f_2^{n+1} .
\end{split}
\end{equation} 
Here, the solution is determined by solving each equation separately. We can formulate a similar theorem, which describes the stability of this scheme:
\begin{thm}
 Splitting scheme (\ref{eq:u_split})--(\ref{eq:p2_split}) 
is unconditionally stable for  $2\theta \geq 1 + \delta$,
where $\delta = \nu_{\max}$ is the maximum eigenvalue of spectral problem (\ref{eq:eigen_b}).
\end{thm} 

\section{Numerical Experiment}

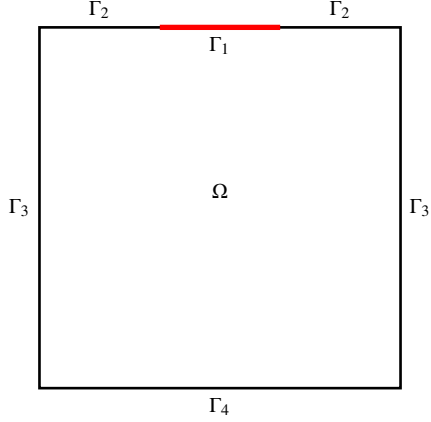
\begin{figure}[!h]
\begin{center}
\begin{tikzpicture}[font=\footnotesize,scale=0.8]
   
\draw [black, line width=1.0] (-3, 0)  rectangle (3, 6) ;
\draw [red, line width=2.0] (-1, 6) -- (1, 6);
\node [above] at (0, 3) {$\Omega$}; 
\node [below] at (0, 6) {$\Gamma_1$}; 
\node [above] at (-2, 6) {$\Gamma_2$}; 
\node [above] at ( 2, 6) {$\Gamma_2$}; 
\node [left]  at (-3, 3) {$\Gamma_3$}; 
\node [right] at ( 3, 3) {$\Gamma_3$};
\node [below] at ( 0, 0) {$\Gamma_4$};  
\end{tikzpicture}
\caption{Computational domain}
\label{fig:domain}
\end{center}
\end{figure}

We consider a two-dimensional double-porosity poroelasticity problem in a unit square domain $\Omega$ subjected to a surface load $\bm g$ as in Fig. \ref{fig:domain}.
At some centered upper part of the domain $\Gamma_1$ the load $\bm g = -\mathrm{sin}(\pi t)\bm n$ is applied in strip of length 0.2 m. The remaining of the top boundary $\Gamma_2$ is traction free. On the vertical boundaries $\Gamma_3$ we set zero horizontal displacement and vertical surface traction. The bottom of the domain $\Gamma_4$ is assumed to be fixed, i.e. the displacement vector is taken as zero.
For pressures, we  prescribe both pore and fracture pressure to be zero at $\Gamma_2$, while the remaining boundaries are impermeable. More precisely, the boundary conditions are given as follows
\[
\begin{split}
&p_1 = 0, \quad p_2 = 0, \quad \bm \sigma \bm n = 0, \quad \bm x \in \Gamma_1, \\
&\pd{p_1}{n} = 0, \quad \pd{p_2}{n} = 0, \quad \bm \sigma  \bm n = \bm g, \quad \bm x \in \Gamma_2, \\
&\pd{p_1}{n} = 0, \quad \pd{p_2}{n} = 0, \quad (\bm \sigma  \bm n) \times \bm n = 0, \quad \bm u \bm n = 0, \, \bm x \in \Gamma_3,  \\
&\pd{p_1}{n} = 0, \quad \pd{p_2}{n} = 0, \quad  \bm u = 0, \quad \bm x \in \Gamma_4.
\end{split}
\]
Similar mathematical model is commonly used to test computational algorithms for numerical solution of poroelasticity problems \cite{boal2011finite,boal2012finite,gaspar2008stabilized}.

To analyze the considered numerical schemes and to identify the dependence of the stability of splitting schemes from the problem parameters we use three sets of input parameters presented in Table \ref{tab:params}. 
For convenience, we vary only values of the $\beta_1$ and $\beta_2$, and other parameters are the same for all sets.

\begin{table}[!h]
\caption{Problem properties}
\label{tab:params}
\begin{center}
\begin{tabular}{ccccc}
\hline 
Parameter 	& Unit 	& Set 1 & Set 2 & Set 3  \\ \hline 
 $\eta$		& Pa$\cdot$s & $0.001$ & $0.001$ & $0.001$  \\
 $\mu$		& MPa & $4.2$ & $4.2$ & $4.2$    \\
 $\lambda$	& MPa & $2.4$  & $2.4$ & $2.4$   \\
 $\beta_1$  	& GPa$^{-1}$ & $54$ & $108$  & $216$ \\
 $\beta_2$  	& GPa$^{-1}$ & $14$ & $24$ & $48$   \\
 $k_1$		& $10^{-15}$ m$^2$ & $6.18$ & $6.18$ & $6.18$   \\
 $k_2$		& $10^{-15}$ m$^2$ & $27,2$ & $27.2$ & $27.2$  \\
 $\alpha_1$	& $-$ & $0.95$ & $0.95$  & $0.95$  \\
 $\alpha_2$ 	& $-$ & $0.12$  & $0.12$ & $0.12$ \\
 $\gamma$	& $10^{-10}$ kg/(m$\cdot$s) & $5$ & $5$ & $5$  \\
\hline 
\end{tabular} 
\end{center}
\end{table}

\begin{figure}[!h]
\begin{center}
\includegraphics[width=0.8\linewidth]{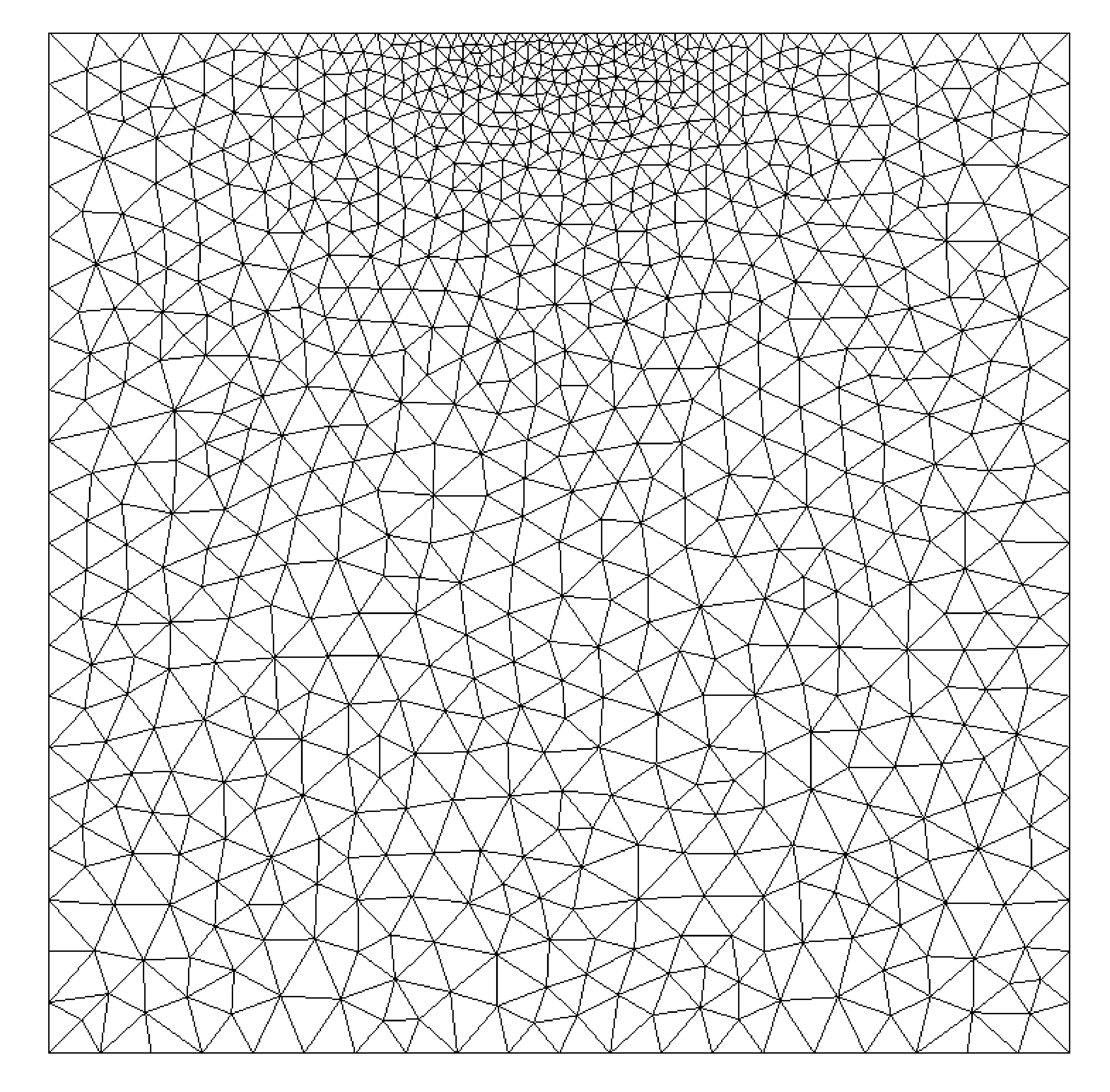}
\caption{Computational mesh 1}
\label{fig:mesh1}
\end{center}
\end{figure}

For the numerical solution we use four computational meshes of different quality with the local refinement in the area of application of the traction. The numbers of vertices, cells, and degrees of freedom (dof) of $\bm u$, $p_1$, $p_2$, and $\bm w = \lbrace \bm u, p_1, p_2 \rbrace$ for each mesh are given in Table \ref{tab:meshes}. Fig \ref{fig:mesh1} shows the coarse mesh 1. Here, the quadratic vector element and the linear scalar element are used for discretization of the displacement and pressures, respectively. Therefore, the number of dofs for the the displacement $\bm u$ is much higher  than the dofs number for the pressures $p_1$ and $p_2$. 

The numerical implementation is performed using the DOLFIN library, which is a part of the FEniCS project for automated solution of differential equations by finite element methods \cite{AlnaesBlechta2015a,LoggMardalEtAl2012a,LoggWells2010a}. The computational meshes are generated using the Gmsh software \cite{geuzaine2009gmsh}. To solve systems of linear equations we use LU  solver provided by PETSc \cite{petsc-web-page}. 

\begin{table}[!h]
\caption{Parameters of meshes}
\label{tab:meshes}
\begin{center}
\begin{tabular}{lrrrr}
\hline 
   			            & mesh 1 	& mesh 2 	& mesh 3		& mesh 4 \\ \hline
Vertices			        & 881		& 3409		& 13398		& 52366 \\
Cells			        & 1654 		& 6604		& 26370		& 103884 \\
Number of $\bm u$ dof	& 6830		& 26842		& 106330 	& 417230 \\
Number of $p_1$   dof	& 881		& 3409		& 13398  	& 52366 \\
Number of $p_2$   dof	& 881		& 3409		& 13398 	 	& 52366 \\
Number of $\bm w$ dof 	& 8592		& 33660	 	& 133126 	& 521962 \\
\hline 
\end{tabular} 
\end{center}
\end{table}

\begin{figure}[!h]
\begin{center}
\includegraphics[width=0.8\linewidth]{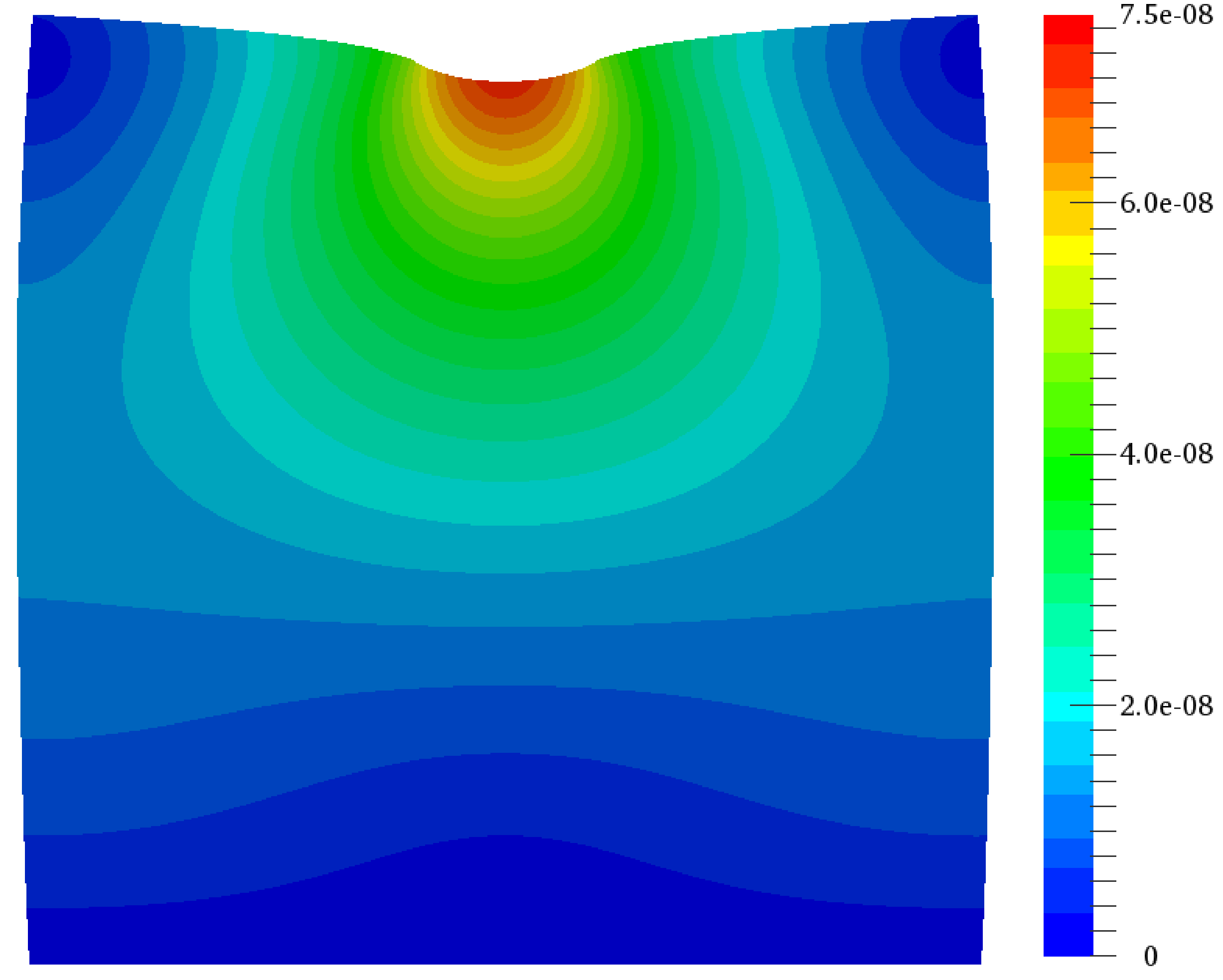}
\caption{Distribution of displacement}
\label{fig:u}
\end{center}
\end{figure}

\begin{figure}[!h]
\begin{center}
\includegraphics[width=0.8\linewidth]{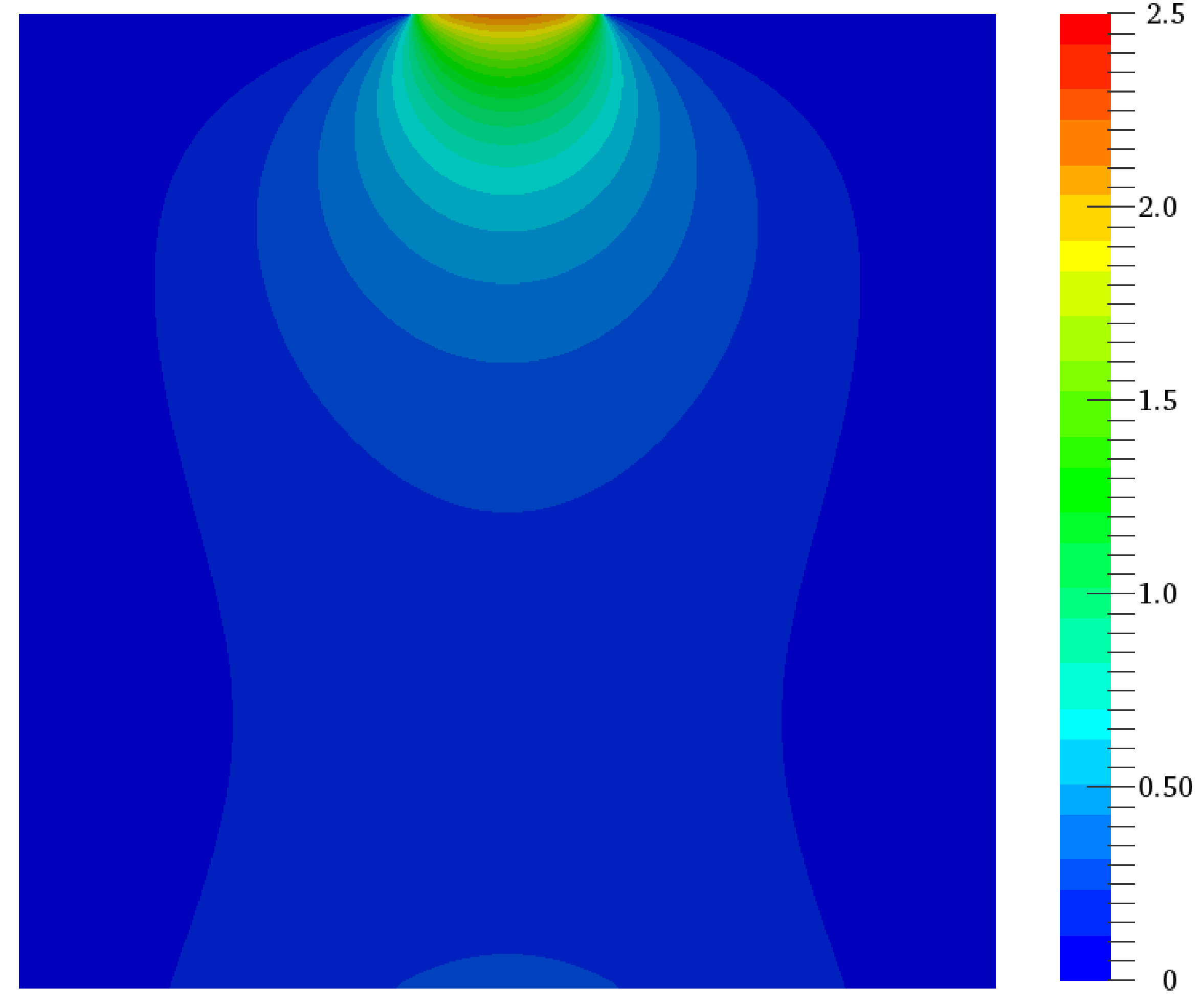}
\caption{Distribution of pressures in pores}
\label{fig:p1}
\end{center}
\end{figure}

\begin{figure}[!h]
\begin{center}
\includegraphics[width=0.8\linewidth]{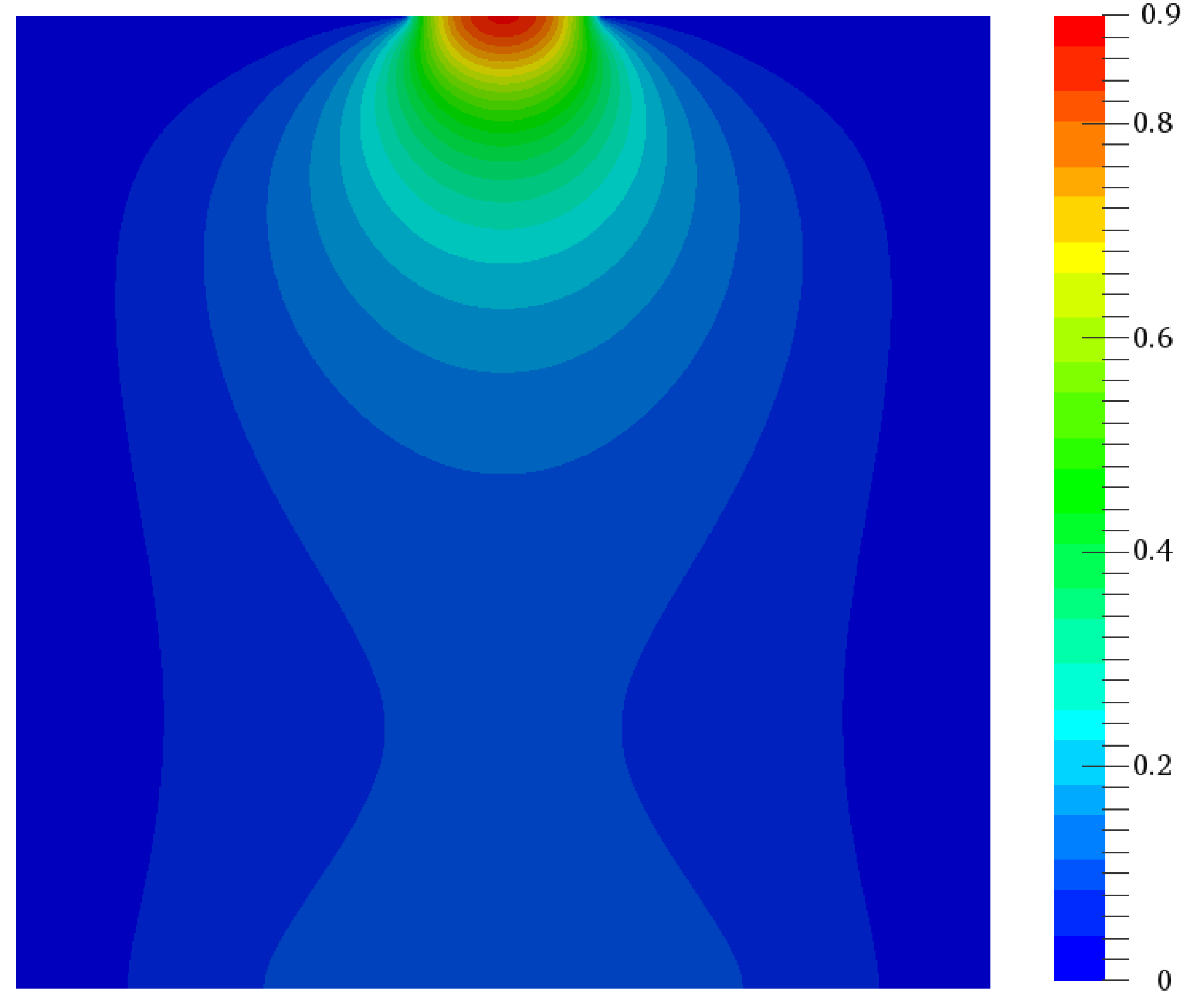}
\caption{Distribution of pressures in fractures}
\label{fig:p2}
\end{center}
\end{figure}

Fig. \ref{fig:u} -- \ref{fig:p2} show the displacement and pressures in pores and fractures, respectively,  at time $t = 0.5$ s, which are obtained using the two-level scheme (\ref{eq:u_coupled})--(\ref{eq:ic_coupled}) with $\theta=1.0$ on the finest mesh 4 and time step $\tau = 0.0025$ s. The displacement is presented on the deformed domain (overstated for visual contrast). We see that the pressure in fractures declines more rapidly than in pores.  
These results will be used as etalon solutions  $\bm u_e$, $p_{1,e}$, and $p_{2,e}$ to evaluate errors as
\[
\varepsilon_{\bm u} = || \bm u_e - \bm u ||, \quad \varepsilon_{p_1} = || p_{1,e} - p_1 ||, \quad \varepsilon_{p_2} = || p_{2,e} - p_2 ||.  
\]

We analyze the dependence of the accuracy of two-level scheme with weights  (\ref{eq:u_coupled})--(\ref{eq:ic_coupled}) from computational parameters: mesh and time step sizes. 
On this stage, we employ only the first set of input parameters  (Table. \ref{tab:params}),
because they do not have an impact on the computational character of two-level scheme. 

\begin{figure}[!h]
\begin{center}
\includegraphics[width=0.8\linewidth]{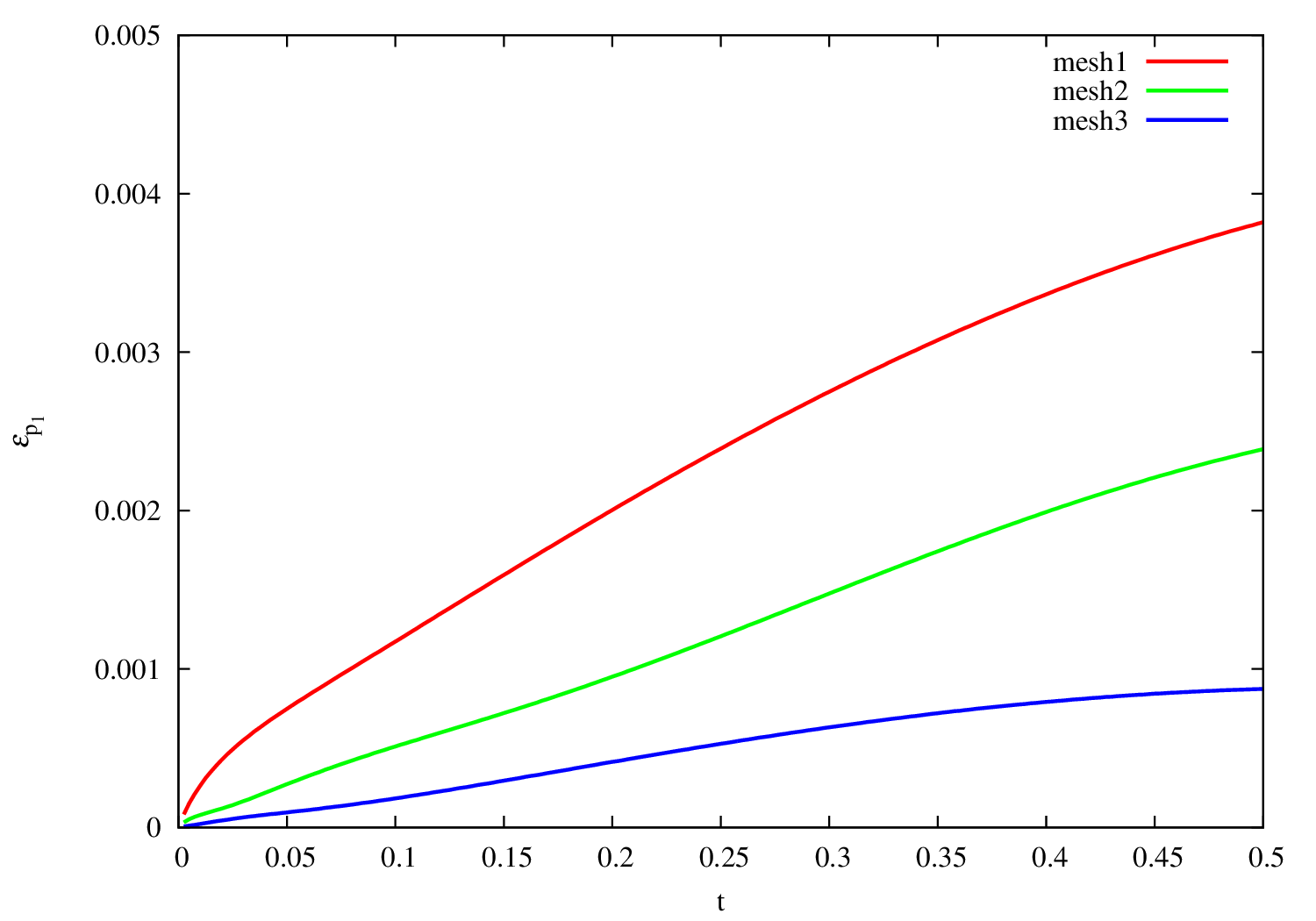}
\caption{Coupled scheme: comparison of the errors $\varepsilon_{p_1}$ for different meshes}
\label{fig:p1_coupled_meshes}
\end{center}
\end{figure}

\begin{figure}[!h]
\begin{center}
\includegraphics[width=0.8\linewidth]{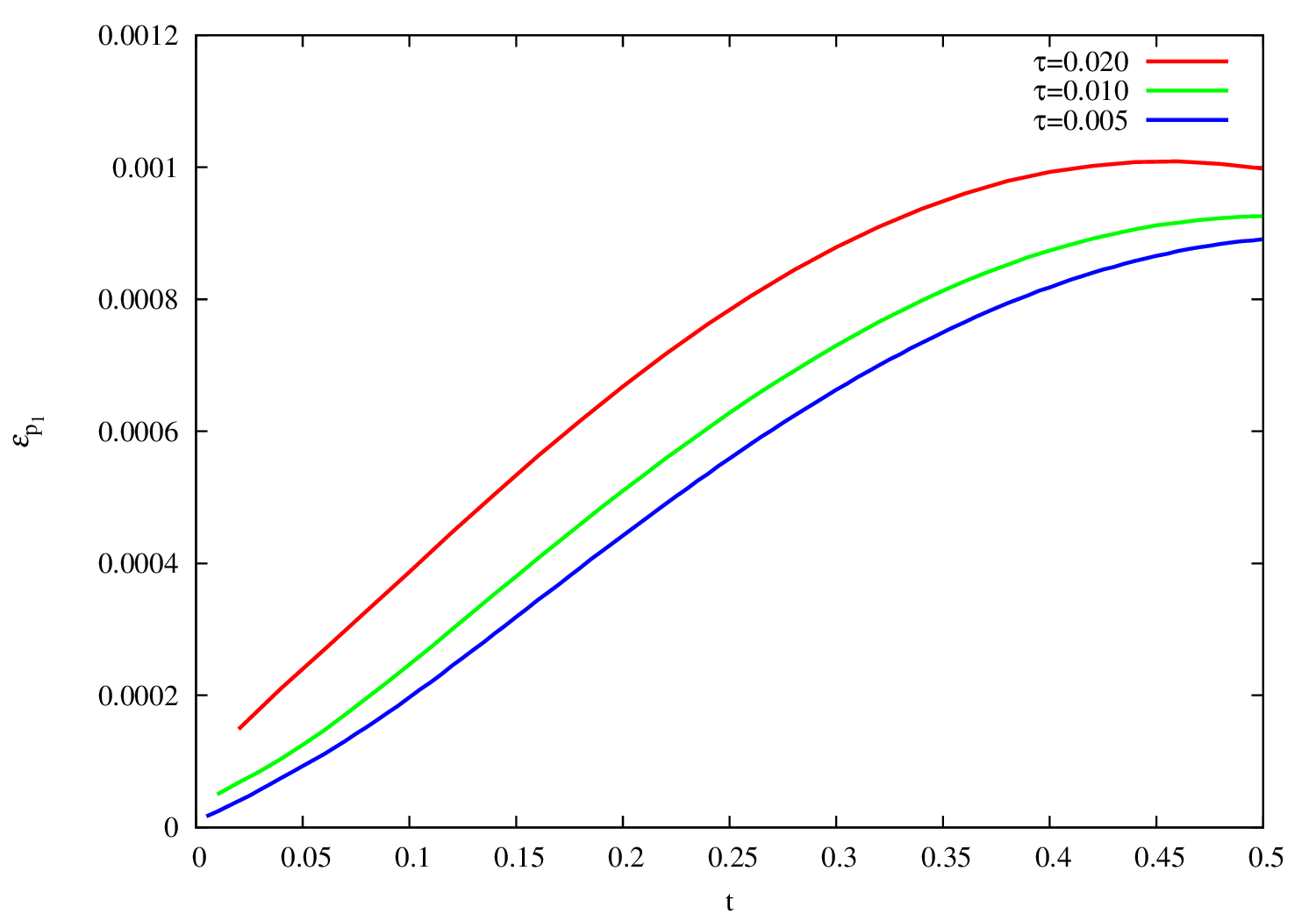}
\caption{Coupled scheme: comparison of the errors $\varepsilon_{p_1}$ for different time steps}
\label{fig:p1_coupled_times}
\end{center}
\end{figure}

Fig. \ref{fig:p1_coupled_meshes} illustrates the dynamics of errors of the pressure in pores $\varepsilon_{p_1}$ for different meshes 1, 2, and 3 with $\tau = 0.0025$ s, while errors for different time steps $\tau = 0.02$, $0.01$, and $0.005$ s on mesh 3 are shown in Fig. \ref{fig:p1_coupled_times}. The weight $\theta=1.0$ is used.
We observe the convergence of solution when  increasing quality of mesh and reducing time step $\tau$. 

Now, we conduct experiments to demonstrate the efficiency of incomplete splitting scheme (\ref{eq:u_split_incomplete})--(\ref{eq:p_split_incomplete}). 
First, using the SLEPs library \cite{Hernandez:2005:SSF} we solve the eigenvalue problem (\ref{eq:eigen_b}) and find the value of $\delta = \nu_{\max}$ and estimate the minimum value of the weight $\theta$ using (\ref{eq:theta_b}), when three-level scheme is unconditionally stable. 
The results are presented in Table \ref{tab:exim_eigen_results}. 
Note that the eigenvalues $\nu_{\max}$ are virtually independent from mesh size and depend only on the problem properties. 

\begin{table}
\caption{Parameter $\delta$ and weight $\theta$ }
\label{tab:exim_eigen_results}
\begin{center}
\begin{tabular}{lccc}
\hline
 & $\quad$ Set 1 $\quad$ & $\quad$ Set 2  $\quad$ & $\quad$ Set 3$\quad$ \\ \hline
 $\delta$ &  2.49 & 1.25 & 0.62 \\
 $\theta$ &  1.75 & 1.12 & 0.81  \\ \hline
\end{tabular}
\end{center}
\end{table}

\begin{figure}
\begin{center}
\includegraphics[width=0.8\linewidth]{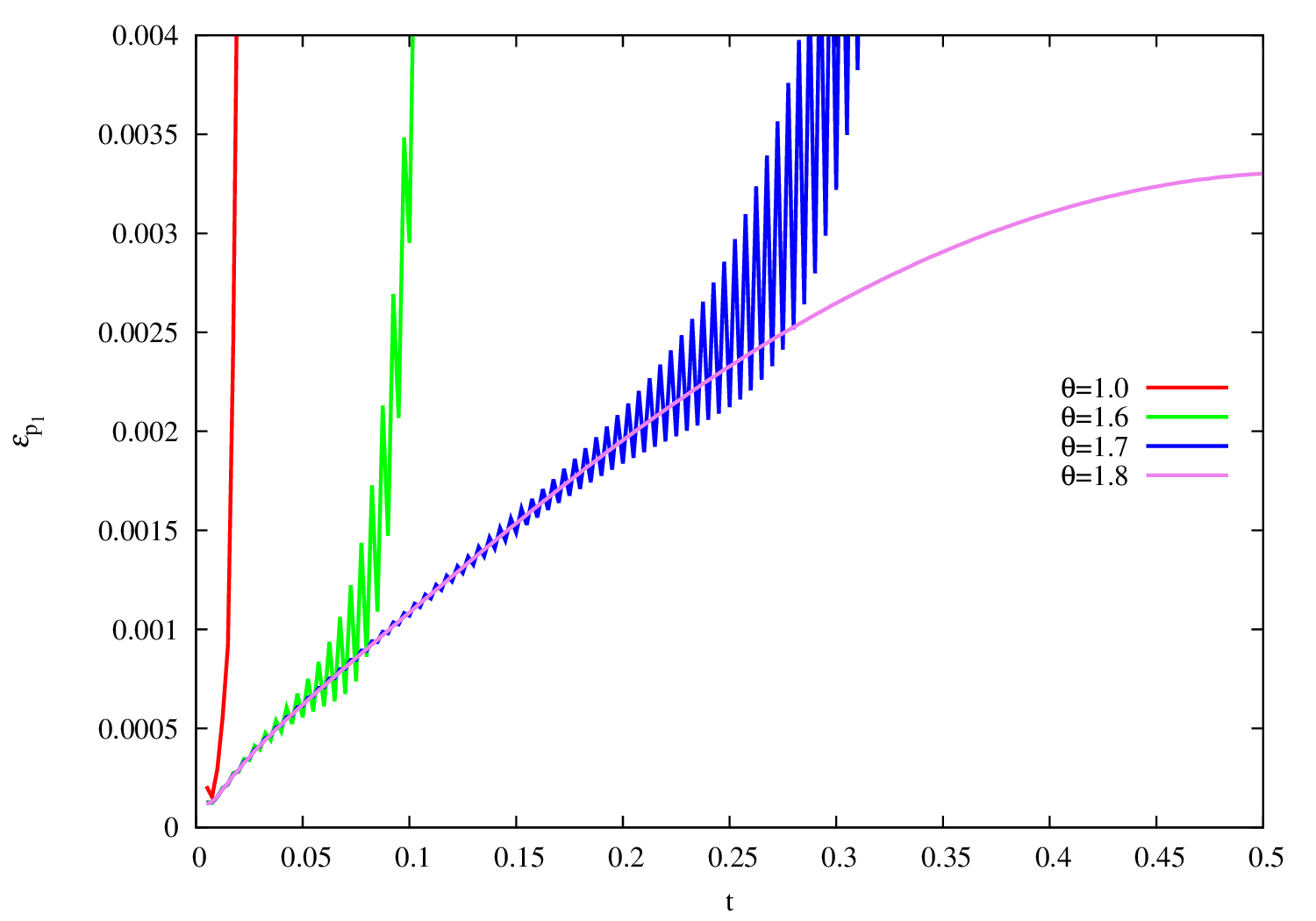}
\end{center}
\caption{Incomplete splitting scheme: comparison of the errors $\varepsilon_{p_1}$  for different $\theta$ and Set 1}
\label{fig:p1_split_test1_thetas}
\end{figure}

\begin{figure}
\begin{center}
\includegraphics[width=0.8\linewidth]{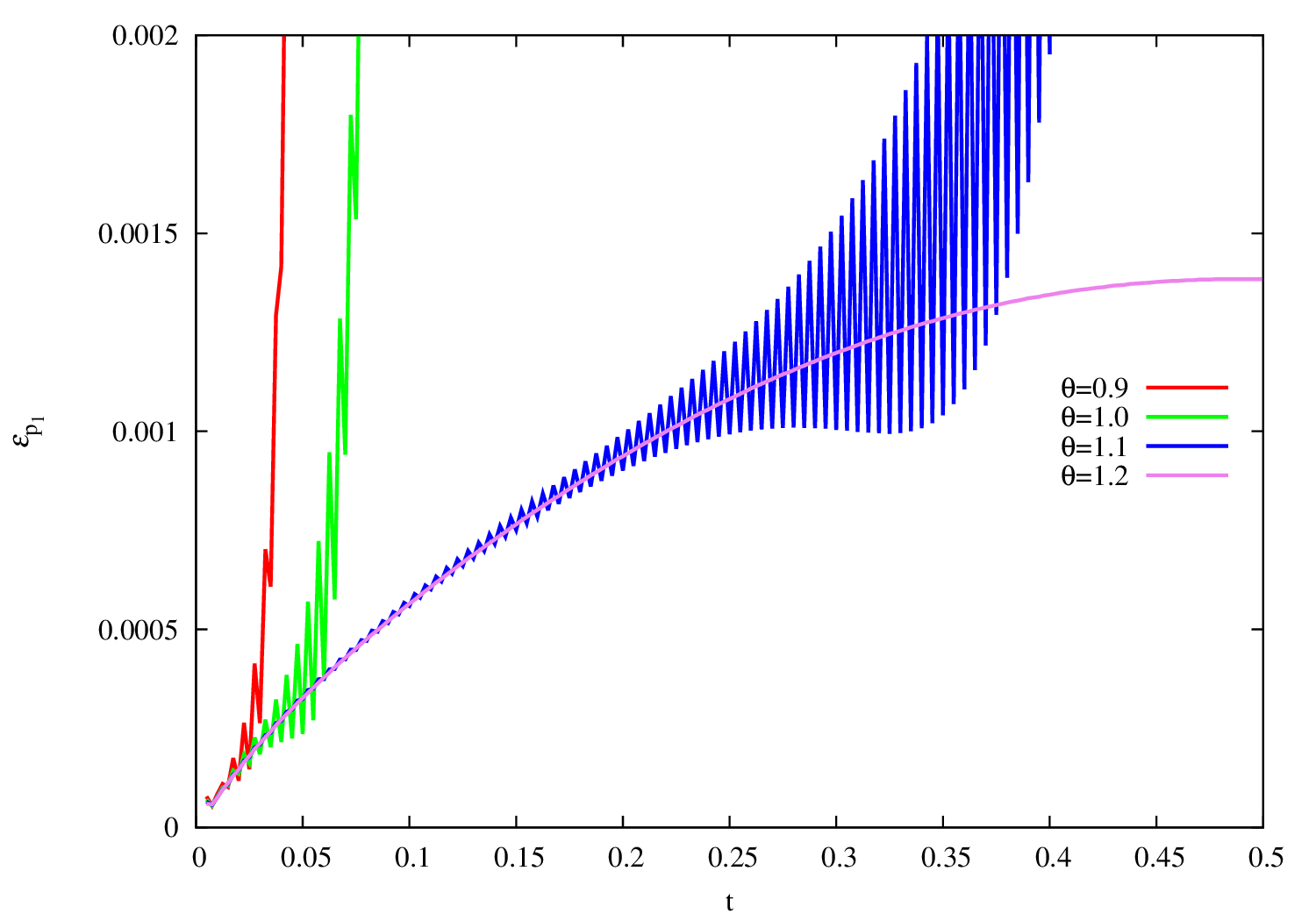}
\end{center}
\caption{Incomplete Splitting scheme. Comparison of the errors $\varepsilon_{p_1}$  for different $\theta$ and Set 2}
\label{fig:p1_split_test2_thetas}
\end{figure}

\begin{figure}
\begin{center}
\includegraphics[width=0.8\linewidth]{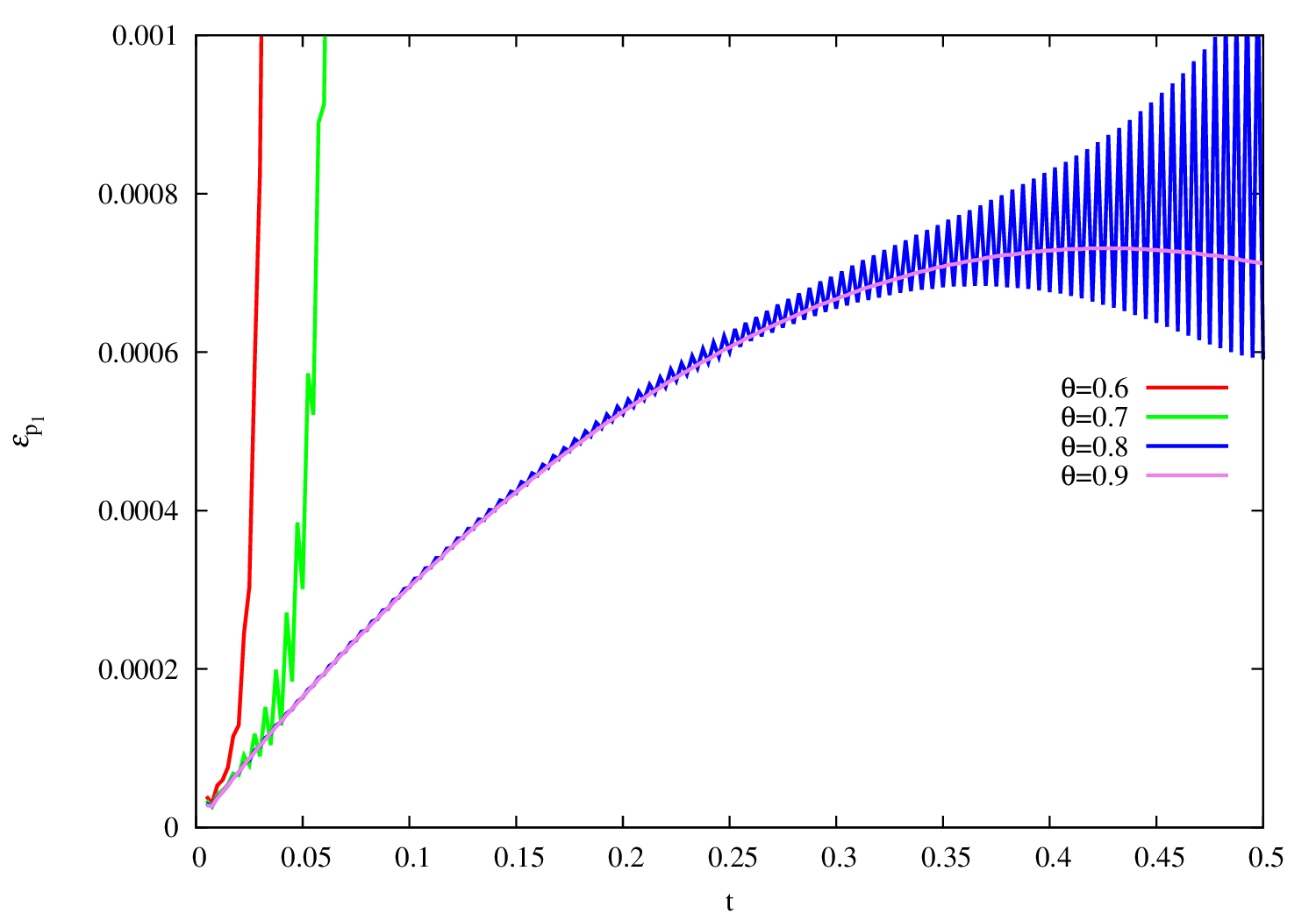}
\end{center}
\caption{Incomplete splitting scheme: comparison of the errors $\varepsilon_{p_1}$  for different $\theta$ and Set 3}
\label{fig:p1_split_test3_thetas}
\end{figure}

In Figs. \ref{fig:p1_split_test1_thetas} -- \ref{fig:p1_split_test3_thetas} we present the dynamics of errors $\varepsilon_{p_1}$  for different values of  $\theta$ for sets 1, 2, and 3, respectively.
In all cases, if the condition (\ref{eq:theta_b}) is not satisfied, then the solution is unstable.
When the value of the weight is taken according to Table \ref{tab:exim_eigen_results}, the solution is regularized. 

\begin{figure}[!h]
\begin{center}
\includegraphics[width=0.8\linewidth]{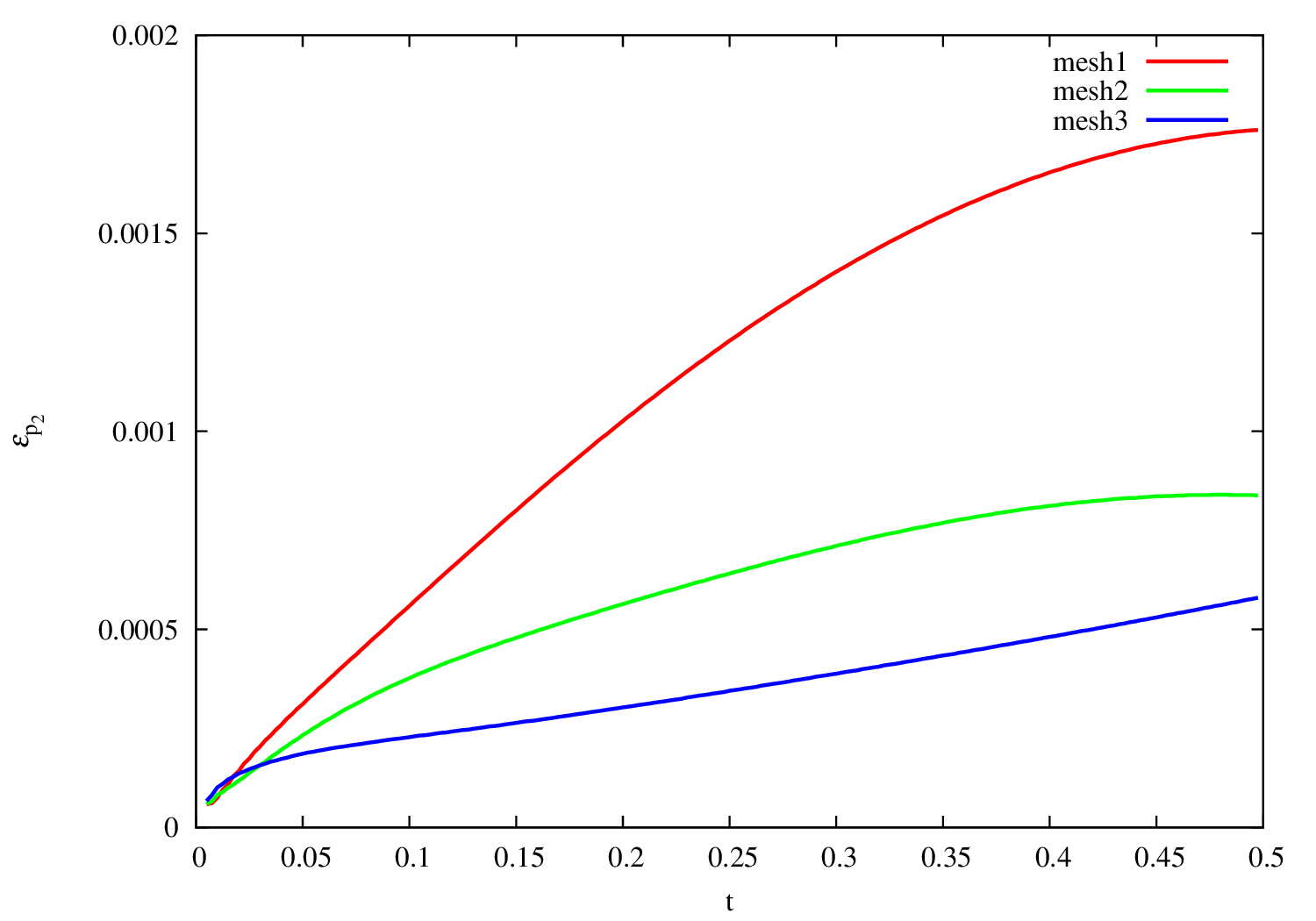}
\caption{Full splitting scheme: comparison of the errors $\varepsilon_{p_2}$ for different meshes}
\label{fig:p2_split_full_meshes}
\end{center}
\end{figure}

\begin{figure}[!h]
\begin{center}
\includegraphics[width=0.8\linewidth]{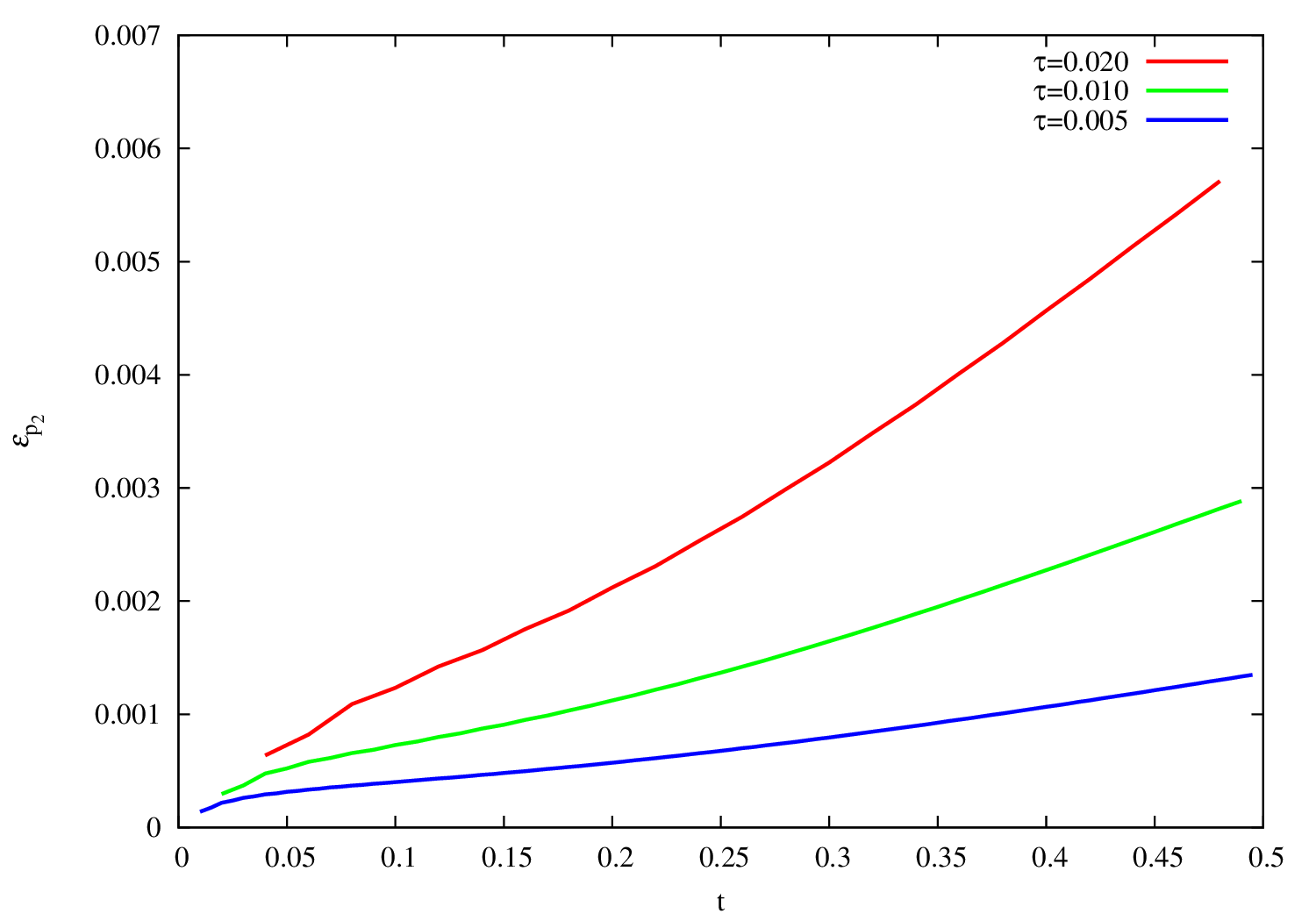}
\caption{Full splitting scheme: comparison of the errors $\varepsilon_{p_2}$ for different time steps}
\label{fig:p2_split_full_times}
\end{center}
\end{figure}

Next, we consider the full splitting scheme (\ref{eq:u_split})--(\ref{eq:p2_split}). 
The dynamics of errors of the pressure in fractures $\varepsilon_{p_2}$ for different meshes and time steps are shown in Figs. \ref{fig:p2_split_full_meshes} and \ref{fig:p2_split_full_times}. Here, we use the first set of input parameters  (Table \ref{tab:params}) and $\theta=1.8$. We also see good convergence of solution. 

Finally, in Table \ref{tab:solve_time} we present the dependence of solve time of coupled scheme (\ref{eq:u_coupled}), (\ref{eq:p_coupled}), incomplete splitting scheme (\ref{eq:u_split_incomplete}), (\ref{eq:p_split_incomplete}), and full splitting scheme (\ref{eq:u_split})--(\ref{eq:p2_split}) from mesh size. Here, the first set of input parameters and $\tau=0.005$ are used.  We see that splitting schemes are significantly faster than coupled scheme. When mesh is small, the solve times of  the incomplete and full splitting schemes differ little from each other. When mesh is big, the full splitting scheme is much faster than incomplete scheme. 

\begin{table}
\caption{Solve time for different meshes and numerical schemes}
\label{tab:solve_time}
\begin{center}
\begin{tabular}{lrrr}
\hline
 & Full splitting & Incomplete  & Coupled  \\
 &   scheme & splitting scheme & scheme  \\ \hline
mesh 1 & 6.894	&	7.045	&	12.684 \\
mesh 2 & 27.578	& 	28.452	&	55.892 \\
mesh 3 & 154.824	&	165.958	&	305.932 \\ 
mesh 4 & 866.452 &	1033.207 &	2488.546 \\
\hline
\end{tabular}
\end{center}
\end{table}

\section{Conclusion}

In this paper, we constructed unconditionally stable three-level splitting schemes with weights for numerical solution of  double-porosity poroelasticity problems.
The analysis was based on the general theory of stability and correctness of operator-difference schemes.  
The finite element method was used approximation in space. 
The efficiency of considered schemes were verified by numerical experiments.

\end{document}